\documentclass[12pt]{article}
 \usepackage{graphicx}

\begin{document}
 \title{Anomalous isotope effects of fulleride superconductors}
 \author{Wei Fan \\
 Key Laboratory of Materials Physics, \\
Institute of Solid State Physics,
 Chinese Academy of Sciences, \\ 230031-Hefei, People's Republic of
 China}
 \date{\today}
 \maketitle
 \begin{abstract}
 The numerical calculations of the standard Eliashberg-Nambu strong coupling theory and a formula of isotope effect derived in this paper provide direct evidences that an-harmonic vibrations of lattice enhance isotope effect with anomalous coefficient $\alpha>1/2$. The results in this paper explain very well the wide distributed $\alpha$ values for the samples with different ratios of substitutions of $^{12}$C by $^{13}$C of fulleride superconductor Rb$_{3}$C$_{60}$. The calculations of isotope effects indicate that the intra-molecule radial modes have more important contributions to superconductivity than the intra-molecule tangential modes with higher phonon frequencies
 \end{abstract}
 \noindent{\it PACS}: 74.20.Fg, 74.70.Wz, 71.20.Tx, 81.05.Tp

 \vspace{2pc}
 \noindent{\it Keywords}: Fulleride superconductor, Anomalous isotope effect, An-harmonic effect, Eliashberg-Nambu Theory

 \section{Introduction}
 The discoveries of isotope effects in metallic superconductors confirm the electron-phonon mechanism as the origin of attractive interaction between electrons. Anomalous isotope effects such as the negative isotope effects with $\alpha<0$ and the enhanced isotope effects with $\alpha>1/2$ are found in many new types of superconductors. Understanding the origin of these anomalous isotope effects is very important to understand the microscopic mechanism of superconductivity.

 Coefficient of isotope effect $0<\alpha<1/2$ is correct for most conventional superconductors. Negative isotope effect $\alpha<0$, where T$_{c}$ increases with mass M, had been found experimentally in conventional superconductors such as PdH~\cite{Skoskiewicz1,Miller1} ($\alpha=-0.25$), organic superconductors~\cite{Schlueter1,Kini1} and $\alpha$-Uranium ($\alpha=-2.0$)~\cite{Fowler1}. The negative isotope effect had also been found in Sr$_{2}$RuO$_{4}$~\cite{Mao1} when $^{16}$O atoms were substituted by $^{18}$O atoms.

 Large positive Oxygen isotope effects $\alpha>1/2$ beyond the BCS value $1/2$ had been found in HTSC material La$_{2-x}$Sr$_{x}$CuO$_{4}$ ($\alpha$=0.75) at doping level x=0.12~\cite{Crawford1} and in fullerides Rb$_{3}$C$_{60}$ ($\alpha>1.0$) with different ratios of substitutions of $^{12}$C with $^{13}$C~\cite{Zakhidov1,Auban-Senzier1,Ebbesen1}. For La$_{2-x}$Sr$_{x}$CuO$_{4}$, near x=0.12 the large coefficient of isotope effect is probably due to the strong an-harmonic vibrations companying with structural transitions. Similarly, the transition from partial substitution to completed substitution of $^{12}$C by $^{13}$C for Rb$_{3}$C$_{60}$ leads to the decrease of $\alpha$ from 1.275 or 1.189~\cite{Zakhidov1,Auban-Senzier1} to 0.30~\cite{Chen1} because the an-harmonic effect is relatively weak at completed substitution. The Eliashberg theory combined with different models of an-harmonic lattice vibrations had been used to qualitatively explain these anomalous isotope effects~\cite{Galbaatar1,Galbaatar2,Inada1}. Beyond Migdal's theorem, the coefficient of isotope effect $\alpha>1/2$ had also been obtained by including non-adiabatic effect~\cite{Grimaldi1}.

 It's very convenient to find a formula for isotope effect which is able to explain anomalous isotope effects such as the enhanced isotope effects with $\alpha>1/2$ and large negative isotope effect. Most of previous formulas of isotope effects don't include the frequency shift due to the isotope substitution. A formula for isotope effect is derived in this paper using McMillan's T$_{c}$ formula, which explicitly includes the frequency shift of phonon and is qualitatively consistent with the more accurate numerical results of Eliashberg-Nambu theory. The formula is similar to a previous formula derived with similar idea~\cite{Crespi1}. In this work, we not only compare the results of the formula with the strong-coupling theory, but also we study in details the meaning of the an-harmonic coefficient $A^{*}$ and its relation with the potential $V(r)$ of the vibrations of atoms around their equilibrium positions in a crystal.  We concentrate on the anomalous isotope effects of fulleride superconductors~\cite{Gunnarsson1} in the present paper.

An important relation is the key to understand isotope effect and written as
 \begin{equation}\label{constraint}
 M\langle\omega^{2}\rangle\lambda=\eta=const
 \end{equation}
 \noindent where the McMillan-Hopfield parameter $\eta$ characterizes the chemical environment of atoms in a material~\cite{Allen1,McMillan1} and is expressed as $\eta=N(0)\langle J^{2} \rangle$, where $N(0)$ is the density of state at Fermi energy, $J$ the matrix element of electron-phonon interaction. Certainly, we assume $\eta$ is constant for isotope substitution or others small structural changes. We will see in section (\ref{isotope_effects}) that, under the constraint, the an-harmonic effect is equivalent to the fact that the parameter $\lambda$ is dependent on $M$ the mass of atom. In a previous work, using electron-phonon mechanism and the constraint~Eq.(\ref{constraint}), we have successfully explained the spatial anti-correlation of the energy gap and the phonon energy of Bi2212 superconductors observed in STM experiments~\cite{Fan1}. In this work, using the constraint combining with McMillan's T$_{c}$ formula, we derive a formula of isotope effect that can explain almost all anomalous isotope effects, especially for fulleride superconductors.

\section{\label{theory} Theory}

 The energy-gap equation of Eliasgberg-Nambu theory in the Matsubara's imaginary energy form is standard~\cite{Allen1}. When temperature is very close to transition temperature T$_{c}$, the Eliashberg equation can be simplified as \begin{eqnarray}\label{energy_gap}
 && \sum_{n=0}^{N}(K_{mn}-\rho\delta_{mn})\bar{\Delta}_{n}=0~~(m \geq 0) \\ \nonumber
 K_{mn}&=&\lambda(m-n)+\lambda(m+n+1)-2\mu^{*}(N) \\ \nonumber
 &-&\delta_{mn}(2m+1+\lambda(0)+2\sum_{l=1}^{m}\lambda(l)),
 \end{eqnarray}
 \noindent where $\bar{\Delta}_{n}=\Delta_{n}/|\omega_{n}|$ is the energy-gap parameter and $\lambda(n)=2\int_{0}^{\infty}d\omega\alpha^{2}F(\omega)\omega/(\omega^{2}+(2\pi nT)^{2})$. The pair-breaking parameter $\rho$ is introduced to form a eigenvalue problem with $\rho$=0 corresponding the physical energy-gap equation. The transition temperature T$_{c}$ is defined as the temperature when the maximum of eigenvalue of kernel matrix $K_{mn}$ crosses to zero and changes its sign. We use about N=200 Matsubara energies to solve above equation.

 The Eliashberg function is expressed as
 \begin{equation}\label{Eliashberg}
 \alpha^{2}F(\omega)=\left\{
 \begin{tabular}{cc} $\frac{c}{(\omega-\Omega_{P})^{2}+(\omega_{2})^{2}}
 -\frac{c}{(\omega_{3})^{2}+(\omega_{2})^{2}}$, & $|\omega-\Omega_{P}|<\omega_{3}$ \\ 0 & others, \end{tabular}\right.
 \end{equation}
 \noindent where $\Omega_{P}$ is the energy  (or frequency) of phonon mode, $\omega_{2}$ the half-width of peak of phonon mode and $\omega_{3}=2\omega_{2}$. We can write the parameter of electron-phonon interaction $\lambda=\lambda(0)=2\int_{0}^{\infty}d\omega\alpha^{2}F(\omega)/\omega$. The moments $\langle\omega^{n}\rangle$ of the distribution function $(2/\lambda)\alpha^{2}F(\omega)/\omega$ are defined as $\langle\omega^{n}\rangle=2/\lambda\int^{\infty}_{0}d\omega\alpha^{2}F(\omega)\omega^{n-1}$. The parameter $\lambda$ characterizes the strength of electron-phonon interaction by $\lambda\propto N(0)\langle J^{2}\rangle/M\langle\omega^{2}\rangle$ in terms of Eq.(\ref{constraint}). The Coulomb pseudo-potential is defined as $\mu_{0}=N(0)U$ and its renormalized value as $\mu^{*}=\mu_{0}/(1+\mu_{0}\ln(E_{C}/\omega_{0}))$, where $U$ is the Coulomb parameter, $E_{C}$ the characteristic energy for electrons such as the Fermi energy or band width, and $\omega_{0}$ the characteristic phonon energy such as the energy cutoff of phonon energy or Debye energy. If $\omega_{2}\ll \Omega_{P}$, Eq.(\ref{constraint}) can be simplified as
 \begin{equation} \label{constraint2} M\Omega_{P}^{2}\lambda=\eta=const, \end{equation}
 \noindent which is obviously correct for Einstein model $\alpha^{2}F(\omega)=(\lambda/2)\Omega_{P}\delta(\omega-\Omega_{P})$.

\section{\label{New_Formula} A formula of coefficient of isotope effect}

 We will derive a formula of isotope effect by including an-harmonic effect. The shift of phonon energy (or frequency) is explicitly included in the formula. For single-frequency like mode, the $\alpha$ values in terms of the formula are qualitatively consistent with the numerical solutions of Eliashberg equation. Most importantly, the formula clearly shows that the an-harmonic vibration of lattice leads to the enhanced isotope effect with $\alpha>1/2$. It's conveniently to define a parameter $A^{*}$ by $\Omega_{P}\propto M^{-(1-A^{*})/2}$ to measure the an-harmonic effect. We can easily see that $A^{*}\neq 0$ represents the an-harmonic effect. The an-harmonic parameter $A^{*}$ can be expressed as
\begin{equation}\label{anharmonic}
A^{*}=1+2M/\Omega_{P}(\delta\Omega_{P}/\delta M).
\end{equation}
Under harmonic approximation $\Omega_{P}\propto M^{-1/2}$, the parameter $\lambda$ is a constant ($\delta\lambda/\delta M=0$) because
$$
 A^{*}=1+2M/\Omega_{P}(\delta\Omega_{P}/\delta
 M)=-M/\lambda(\delta\lambda/\delta M)=0
$$
\noindent by using the relation $M\Omega_{P}^{2}\lambda=const$.

 We start from the McMillan formula of transition temperature T$_{c}$ of superconductor
 \begin{equation}\label{McMi} T_{c}=\frac{\Theta_{D}}{1.45}\exp[-\frac{1.04(1+\lambda)}{\lambda-\mu^{*}(1+0.62\lambda)}]. \end{equation}
 \noindent where $\lambda$ is the strength of electron-phonon interaction. The coefficient $\alpha$ of isotope effect defined by T$_{c}\propto M^{-\alpha}$ can be obtained from above McMillan formula by the direct mass-dependent from $\Theta_{D}\propto M^{-1/2}$ and implicit mass-dependent from $\mu^{*}$ by $\omega_{0}\propto M^{-1/2}$. If an-harmonic effect is included then $\Theta_{D},\omega_{0} \propto M^{-(1-A^{*})/2}$. In this work we consider additionally mass-dependent from $\lambda$ by the well known constraint $M\Omega^{2}_{P}\lambda=\eta$, we obtain a formula of isotope effect which is expressed as
 \begin{equation}\label{New_Eq} \alpha=\frac{1}{2}-\frac{1.04(1+\lambda)(1+0.62\lambda)(\mu^{*})^{2}}{2[\lambda-\mu^{*}(1+0.62\lambda)]^{2}}
 +A^{*}T(\lambda,\mu^{*})
 \end{equation}
 \noindent where $A^{*}$ is defined in Eq.(\ref{anharmonic}) and the function $T(\lambda,\mu^{*})$ is expressed as \begin{eqnarray} \label{Talpha} T(\lambda,\mu^{*})=
 \frac{1.04\lambda (1+0.38\mu^{*})}{[\lambda-\mu^{*}(1+0.62\lambda)]^{2}}
 \end{eqnarray} or
 \begin{eqnarray} \label{Talphb}
 T(\lambda,\mu^{*})&=&\frac{\lambda(2.08-\lambda)+\lambda
 (2.7904+1.24\lambda)\mu^{*}}{2[\lambda-\mu^{*}(1+0.62\lambda)]^{2}}
 \\ \nonumber &+&\frac{(0.04+0.42\lambda)(1+0.62\lambda)(\mu^{*})^{2}}{2[\lambda-\mu^{*}(1+0.62\lambda)]^{2}},
 \end{eqnarray}
 \noindent dependent on whether $\Theta_{D},\omega_{0}\propto M^{-1/2}$ for Eq.(\ref{Talpha}) or $\Theta_{D},\omega_{0}\propto M^{-(1-A^{*})/2}$ for Eq.(\ref{Talphb}). We notice that, for the Eq.(\ref{Talpha}), the an-harmonic effect enters into the coefficient $\alpha$ only by the M-dependent $\lambda$, however for the Eq.(\ref{Talphb}) not only by M-dependent $\lambda$ but also by the $\Theta_{D}$  and $\omega_{0}$. It's very important that $T(\lambda,\mu^{*})>0$ for reasonable values $0<\lambda<2<2.08$, so the sign of the third term of Eq.(\ref{New_Eq}) is determined by the an-harmonic parameter $A^{*}$. It's obviously that $A^{*}=0$ under harmonic approximation $\Omega_{P}\propto M^{-1/2}$  and the third term in the Eq.(\ref{New_Eq}) is also zero. The derivative $\delta\Omega_{P}/\delta M$ used in the calculation of  an-harmonic parameter $A^{*}$ is approximately obtained from the experimental energy (or frequency) shift of phonon. The first two terms give $\alpha<1/2$. The properties of the third term to isotope effect is determined by the sign of the an-harmonic parameter $A^{*}$. Generally, $\Omega_{P}$ decreases with increasing mass M, thus $2(M/\Omega_{P})\delta\Omega_{P}/\delta M <0$. If $2|(M/\Omega_{P})\delta\Omega_{P}/\delta M|<1$, $A^{*}>0$. So we can get $\alpha>1/2$ if the third term has larger absolute value than the second term.

\section{\label{isotope_effects} The anomalous isotope effects with $\alpha>1/2$}

 In many literatures of A$_{n}$C$_{60}$, the average mass $m$ of C$_{60}$ is used in the definition of coefficient of isotope effect by $\tilde{\alpha}=-(m/T_{c})\delta T_{c}/\delta m$. If the ratio of substitution is $p$, the average mass of C$_{60}$ molecule $m=60[M_{12}(1-p)+M_{13}p]$, so $\delta m=m-60M_{12}=60p\delta M$ with $\delta M=M_{13}-M_{12}$. The relation $\tilde{\alpha}=\alpha /p$ connects the parameter $\tilde{\alpha}$ with the usual definition of $\alpha=-(M/T_{c})(\delta T_{c}/\delta M)$. The usual coefficient of isotope effect $\alpha=p\tilde{\alpha}=1.275$ is obtained with $\tilde{\alpha}$=2.125 and $p$=0.60~\cite{Zakhidov1}, $\alpha=1.189$ with $\tilde{\alpha}$=1.45 and $p$=0.82~\cite{Auban-Senzier1}, $\alpha=0.462$ with $\tilde{\alpha}$=1.4 and $p$=0.33~\cite{Ebbesen1}. For completed substitution with $p$=1.0, $\alpha=\tilde{\alpha}$=0.30~\cite{Chen1}.

 The relation $M\langle\omega^{2}\rangle\lambda=\eta=const$ services as a constraint to determine the parameters $\lambda$ and $\Omega_{P}$ in numerical calculations. The Coulomb pseudo-potential $\mu^{*}$ has to change in isotope substitution because $\mu^{*}$ is dependent on the cutoff of phonon energy $\omega_{0}\propto M^{-1/2}$ or $\propto M^{-(1-A^{*})/2}$ in an-harmonic approximation. In harmonic approximation, $\delta\mu^{*}=-(\mu^{*})^{2}\delta M/2M$, the an-harmonic effect is included in calculations only by the M-dependent $\lambda$. The an-harmonic effect can be realized by shifting phonon energy $\delta\Omega_{P}$ to make $A^{*}=1+2M/\Omega_{P}(\delta\Omega_{P}/\delta M) \neq 0$.

 The intra-molecule radial mode, which is about $\Omega_{P}$=65 meV or 525 cm$^{-1}$ from infrared spectrum~\cite{Ebbesen1,Zakhidov1} and the intra-molecule tangential modes around 1400$^{-1}$ or 174 meV~\cite{Gunnarsson1} have strong intensity. However we concentrate attentions on the intra-molecule radial mode with energy $\Omega_{P}$=65 meV and half-width $\omega_{2}$=8 meV. The Coulomb parameter $\mu_{0}=UN(0)$ and the corresponding renormalized Coulomb parameter $\mu^{*}=\mu_{0}/[1+\mu_{0}\ln(E_{C}/\omega_{0})]$ can be estimated from ab-initio density functional theory (DFT) based on pseudo-potential method using atomic orbital basis functions\cite{Soler1}. In the DFT calculation, the super-cell includes one C$_{60}$ molecule and three Rb atoms. The effect of orientation of different C$_{60}$ molecules is ignored. The parameter $U$ is the charge energy defined as $\delta^{2}E_{tot}/\delta n^{2}=E(n+1)+E(n-1)-2E(n)$, $E(n)$ the total energy of electric-neutral system of n valence electrons, $E(n+1)$ and $E(n-1)$ the total energies with one negative and one positive charge respectively. We choose the possible valence-electron configuration $4p^{6}5s^{1}$ for Rb atoms and $2s^{2}2p^{2}$for Carbon atoms. The core electrons are presented by Troulier-Martins pseudo-potentials. The electrons in semicore state 4p of Rb atoms having already treated as valence electrons have the single-$\zeta$ basis set and all others valence electrons for all atoms have the split valence double-$\zeta$ plus polarized basis sets. We use $\Gamma$ point sampling the first Brillouin zone. The exchange-correlation potential is GGA Perdew-Burke-Ernzerhof type\cite{Perdew1} and the spin-polarization effects are included in the self-consistent calculations. The results in table(\ref{TabMu}) show that the Coulomb parameter $\mu_{0}$ is equal to 4.374 and the renormalized Coulomb parameters $\mu^{*}$ is 0.127 if the maximum of phonon energy $\omega_{0}$ is 150meV and $E_{C}$=15 eV when all valence electrons are included. The common value $\mu^{*}$=0.10 is close 0.127 obtained in this work. We have known the Coulomb parameter $\mu^{*}$ and the phonon energy $\Omega_{P}$, the parameter $\lambda$ of electron-phonon interaction can be defined by the experimental transition temperature. From table(\ref{TabMu}), we can see that the parameters $\mu^{*}$ have small influence on isotope effects. Below we present the calculations of coefficient $\alpha$ of isotope effects with $\mu^{*}$=0.1.

\begin{table}
\caption{\label{TabMu} The Coulomb parameter $\mu^{*}$ is calculated when we choose valence-electron configuration (a) $4p^{6}5s^{1}$ for Rb atom and $2s^{2}2p^{2}$for Carbon atom in DFT calculations. The case (b) is corresponding to the general value $\mu^{*}$=0.10. The coefficients $\alpha$ of isotope effects are calculated using the phonon-energy shifts from 65 meV to 62.3 meV after the isotope substitution (A$^{*}$=0.003).}
 \begin{tabular}{lllllllll}
 \hline     &U(eV)&N(0)(1/eV)&UN(0)&$\ln(E_{F}/\omega_{0})$&$\mu^{*}$&$\lambda$&T$_{c}$(K)&$\alpha$ \\
 \hline
 (a) &1.991&2.197&4.374&6.9078&0.127& 0.700& 29.544& 0.342 \\
 (b)                      &&&&&0.100& 0.670& 29.820& 0.358 \\

 \hline
\end{tabular}
\end{table}

\begin{table}
\caption{\label{TabCm} The simulation parameters and the results with different an-harmonic parameters $A^{*}$. The coefficients $\alpha$ are calculated based on the numerical solutions of Eliashberg equation and $\alpha '$ from the formula Eq.(\ref{New_Eq})
and Eq.(\ref{Talpha}) derived in this paper.}
\begin{tabular}{cccccccc}
\hline
 $^{13}$C&M (u)&$\Omega_{P}$ (meV)&$\lambda$&$T_{c}$ (K) &$A^{*}$&$\alpha$&$\alpha'$ \\
 \hline
  (1) &13&64.3&0.632&25.5&~0.742&~1.929$\pm$0.10&~2.377 \\
  (2) &13&63.4&0.650&27.2&~0.409&~1.207$\pm$0.09&~1.516 \\
  (3) &13&62.3&0.674&29.0&~0.003&~0.358$\pm$0.09&~0.464 \\
 \hline
  (4) &13&62.2&0.676&29.2&-0.034&~0.282$\pm$0.09&~0.368 \\
  (5) &13&62.0&0.680&29.5&-0.108&~0.132$\pm$0.08&~0.177 \\
  (6) &13&61.8&0.685&29.8&-0.182&~-0.02$\pm$0.08&-0.014 \\
 \hline
  (7) &13&61.6&0.689&30.3&-0.255&-0.165$\pm$0.08&-0.205 \\
  (8) &13&61.2&0.698&30.9&-0.403&-0.457$\pm$0.08&-0.588 \\
  (9) &13&60.0&0.727&33.2&-0.846&-1.305$\pm$0.08&-1.736 \\
  \hline
  $^{12}$C &12&65.0&0.670&29.82 \\
 \hline
\end{tabular}
\end{table}

 The parameter of electron-phonon interaction $\lambda$=0.67 is defined by the experimental transition temperature T$_{c}$ about 29.5(K) using parameters $\mu^{*}$=0.1 and $\Omega_{P}$=65 meV. We calculate the $\eta=\eta_{12}$ for $^{12}$C. The phonon energies after $^{13}$C substitutions are unknown because the phonon energies are dependent on the ratios of $^{13}$C substitutions. We choose the nine possible energies for phonons in table~(\ref{TabCm}), which are all smaller than $\Omega_{P}$=65 meV because the energies (or frequency) decrease with increasing mass of atoms. We also assume that the parameters $\lambda$ of electron-phonon interaction alter after isotope substitutions. The new Eliashberg function $\alpha^{2}F(\omega)$ after isotope substitution by $^{13}$C is obtained from the old one before isotope substitution by simply shifting energy of phonon and scaling it to satisfy the constraint $M\langle\omega^{2}\rangle\lambda=\eta_{13}=\eta_{12}$. Based on the new Eliashberg function $\alpha^{2}F(\omega)$, we can calculate the T$_{c}$ and $\lambda$ after the isotope substitution. The new Coulomb pseudo-potential after isotope substitution $\mu^{*}$=0.099583 is obtained from the formula $\delta\mu^{*}=-(\mu^{*})^{2}\delta M/2M$. The nine possible values of $\lambda$ corresponding to nine possible energies of phonon are collected in table~(\ref{TabCm}). The transition temperatures T$_{c}$ are obtained by solving Eliashberg equation Eq.(\ref{energy_gap}). The coefficients $\alpha$ are easily calculated in terms of the transition temperatures T$_{c}$ after and before isotope substitutions by $\alpha=-\ln(^{13}T_{c}/^{12}T_{c})/\ln(M_{13}/M_{12})$. The an-harmonic parameters $A^{*}$ are calculated in terms of the shifts of phonon energies.

 For the case (2) in table~(\ref{TabCm}), we get $\alpha=1.207$ which is very close to the already known maximum value 1.275 in experiments with uncompleted substitution ($\tilde{\alpha}$=2-2.25, $p$=0.60)~\cite{Zakhidov1}. For sample with completed substitution with $^{13}$C, the value of $\alpha$ decreases to $\sim$0.30~\cite{Chen1}. To explain the interesting results, we assume the increase of ratio of substitution makes the distribution of $^{13}$C more homogenous and the an-harmonic effect become weak with small $A^{*}$. If the phonon energy shifts to 62.3 meV with smaller an-harmonic parameter $A^{*}$=0.003, the value of $\alpha$=0.358 is close to the experimental values from 0.30 to 0.37 for the samples with completed substitution.

\begin{figure}\begin{center}\includegraphics[width=0.55\textwidth]{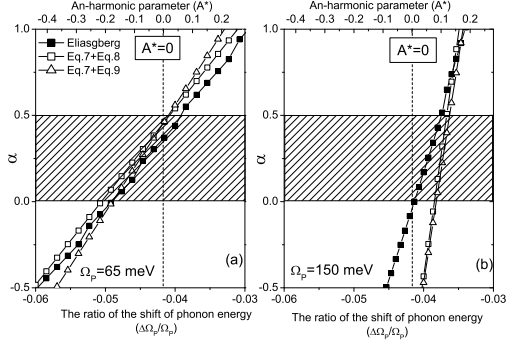}
\caption{\label{fig1} The comparison of coefficients of isotope effect having obtained from numerical calculations of Eliashberg equation and from Eq.(\ref{New_Eq}) for radial intra-molecule mode $\Omega_{P}$=65 meV (a) and the tangential intra-molecule mode $\Omega_{P}$=150 meV (b)}\end{center}\end{figure}

 For negative an-harmonic parameter$A^{*}$=-0.182, we find $\alpha$=0.02 close to zero. After having made $A^{*}$ more negative further, at $A^{*}$=-0.403, we get the negative coefficient $\alpha$=-0.457. However, the negative isotope effect wasn't found in fulleride superconductors. From above calculations we can see that the sign of $A^{*}$ determines the sign of $\alpha$ at relative larger absolute values of $A^{*}$. If the absolute values of an-harmonic parameters are not too large $0.07>A^{*}>-0.21$ or an-harmonic effect is weak, the coefficient $\alpha$ are from 0 to 0.5 within the range of general values. The above calculations explain the wide distributed values of coefficients for Rb$_{3}$C$_{60}$ in experiments. This is because the an-harmonic parameters are dependent on the ratios of substitutions.

 From table~(\ref{TabCm}), we can see that the values obtained from numerical calculations are very close to the values from Eq.(\ref{New_Eq}) with Eq.(\ref{Talpha}).  The values of $\alpha$ and $\alpha$' have the same sign on the same rows. It's very important that if $\alpha>1/2$, $A^{*}>0$ must be satisfied. Thus, the enhanced coefficients $\alpha$ larger than $1/2$ for Rb$_{3}$C$_{60}$ in uncompleted substitution samples mean that the strong an-harmonic effects. If the an-harmonic effect becomes weak as approaching the completed substitution, the value of $\alpha$ will decrease and reach to smaller value about 0.358 at $A^{*}=0.003$.

 Fig. \ref{fig1}(a) shows the results of the numerical solutions of Eliashberg equation and the formula Eq.(\ref{New_Eq}). We can see that Eq.(\ref{New_Eq}) with Eq.(\ref{Talpha}) is more close to the numerical solution than using the Eq.(\ref{Talphb}) although for Eq.(\ref{Talphb}) more completed an-harmonic effects are included. We have preformed the same calculations for intra-molecule tangential mode with energy $\Omega_{P}$=150 meV, $\mu^{*}$=0.285, $\lambda$=0.67 and T$_{c}$=29.5(K). From Fig.\ref{fig1}(b), there is larger slope of $\alpha$-$\delta\Omega_{P}/\Omega_{P}$ curve compared with the intra-molecule radial mode with energy $\Omega_{P}$=65 meV. The coefficient of isotope effect of the intra-molecular radial mode $\alpha$=0.358 at $A^{*}\simeq$0.0 is more close to experiments at completed substitutions $\alpha\sim 0.30-0.37$. So it is more correlated with superconductivity of fulleride than the intra-molecule tangential mode.

\section{An-harmonic parameter $A^{*}$ and models of lattice vibrations}

\begin{figure}\begin{center}\includegraphics[width=0.60\textwidth]{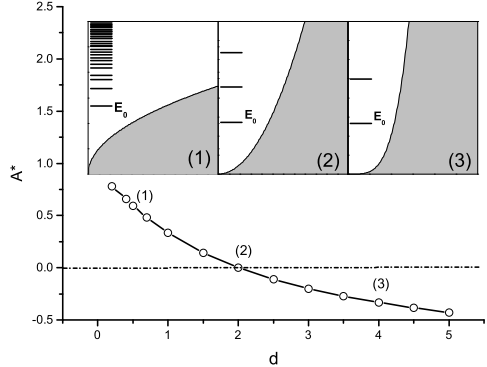}
\caption{\label{fig2} An-harmonic coefficients $A^{*}$ for the models of interaction with form $V(r)=hr^{d}$. The inserted figures (1),(2) and (3) illustrate the shapes of interacting potentials and the quantized energy levels for $h$=0.01 and $d$=0.5, 2.0 and 4.0 respectively. The maximum of quantum number of angular momentum is up to $l_{max}$=8}
\end{center}\end{figure}

In this paper, we introduce an important parameter, the an-harmonic parameter $A^{*}$. To make parameter $A^{*}$ more realistic, we study the models of lattice vibrations in a crystal. The atoms in crystal move around their equilibrium positions. The change of potential energy of an atom away from its equilibrium position is written as $V(r)=hr^{d}$ with radially local displacement $r$. The harmonic vibration is corresponding $d=2$. We had numerically solved the Schr\"odinger equation $[-\hbar^{2}\nabla^{2}/2M+V(r)]\psi(\vec{r})=E\psi(\vec{r})$ under isotropic approximation, especially, the one-dimensional radial part converts into difference equation.  We calculated the relation $E_{0}(M)$ between energy of ground state and mass of atom for a series of discrete values of $M$. It's well known that for harmonic approximation $E_{0}(M)\propto 1/M^{0.5}$. The effective an-harmonic parameters $A^{*}$ for interaction with the form $V(r)=hr^{d}$ are obtained by fitting the discrete $E_{0}(M)$ functions with function $\tilde{E}_{0}(M)\propto 1/M^{(1-A^{*})/2}$. From the Fig.\ref{fig2}, if $d<2$ we can obtain $A^{*}>0$.  So the enhanced isotope effects with $\alpha>0.5$ are hopefully found based on the formula  Eq.(\ref{New_Eq}) obtained in this work. If $d>2$ so $A^{*}<0$, we can get the normal isotope effects with $\alpha<0.5$ and the negative isotope effects with $\alpha<0$. We can see that the simple model $V(r)=hr^{d}$ is suitable for the isotope effects of fulleride Rb$_{3}$C$_{60}$ for different ratios of substitutions because the an-harmonic parameters are within the range to obtain experimental values of $\alpha$.

\section{\label{conclusion} Discussion and Summary}

 The Figure 1(b) shows clearly that the formula Eq.(\ref{New_Eq}) isn't good approximation to more accurate numerical solution for high-energy mode. The reason is probably that McMillan's formula isn't correct in some regions of parameter space. The McMillan's formula is good approximation when the parameter $\lambda$ of electron-phonon interaction is not too large ($\lambda<1$) and the phonon energy $\Omega_{P}$ is not too high.   The inter-molecule phonon modes such as the vibration between $C_{60}$ molecules and the between alkali-metal atoms and $C_{60}$ molecules are ignored in this work because their energies generally smaller than 12 meV. To obtain $T_{c}$=30K, the parameters $\lambda$ of electron-phonon interaction are at least 3.0 so the instability of lattice will destroy superconductivity. However, these inter-molecule modes still have significant influence on the properties of fulleride superconductors such as the differences of isotope effects for different substitution configurations $Rb_{3}[(^{13}C_{60})_{x}(^{12}C_{60})_{1-x}]$ and $Rb_{3}(^{13}C_{x}^{12}C_{1-x})_{60}$~\cite{Chen1}.

 In summary, the coefficients $\alpha$ of isotope effects obtained in this paper are very close to the values in experiments or within the range of experimental values. The reductions of $\alpha$ with increasing the substitution ratios of $^{13}$C are due to the reductions of an-harmonic effects of lattice vibrations when the ratios of substitution tend to 100\%. The enhanced coefficients of isotope effects with $\alpha>1/2$ generally happen at the intermediate stage of transition from one phase to another. Finally, the formula Eq.(\ref{New_Eq}) and the numerical methods used in this paper are also suitable to study isotope effects of other superconductors if the electron-phonon interaction is the pairing mechanism for electrons.

 The author benefits from the discussions with members within CMS Group at ISSP-CAS, especially from Dr. Xu Yong, Huang Ling-Feng, Li Long-Long and Li Yan-Ling. This work is supported by Director Grants of Hefei Institutes of Physical Sciences, Knowledge Innovation Program of Chinese Academy of Sciences and National Science Foundation of China.

\end{document}